\documentclass[aip,apl,reprint]{revtex4-1}
\usepackage{graphicx}

\begin{document}

\title{Tuning domain wall velocity with Dzyaloshinskii-Moriya interaction}

\author{Fernando Ajejas}
\affiliation{Universidad Autonoma de Madrid, Madrid, Spain}
\affiliation{IMDEA Nanoscience, Campus de Cantoblanco, 28049 Madrid, Spain}
\author{Viola K\v{r}i\v{z}\'{a}kov\'{a}}
\author{Dayane de Souza Chaves}
\author{Jan Vogel}
\affiliation{Univ.~Grenoble Alpes, CNRS, Institut N\'eel, F-38000 Grenoble, France}
\author{Paolo Perna}
\author{Ruben Guerrero}
\author{Adrian Gudin}
\affiliation{IMDEA Nanoscience, Campus de Cantoblanco, 28049 Madrid, Spain}
\author{Julio Camarero}
\affiliation{Universidad Autonoma de Madrid, Madrid, Spain}
\affiliation{IMDEA Nanoscience, Campus de Cantoblanco, 28049 Madrid, Spain}
\author{Stefania Pizzini}
\affiliation{Univ.~Grenoble Alpes, CNRS, Institut N\'eel, F-38000 Grenoble, France}
\email[]{stefania.pizzini@neel.cnrs.fr}


\date{\today}

\begin{abstract}
We have studied a series of Pt/Co/M epitaxial trilayers, in which Co is sandwiched between Pt and a non magnetic layer M (Pt,~Ir,~Cu,~Al). Using polar magneto-optical Kerr microscopy, we show that the field-induced domain wall speeds are strongly dependent on the nature of the top layer, they increase going from M=Pt to lighter top metallic overlayers, and can reach several 100~m/s for Pt/Co/Al. The DW dynamics is consistent with the presence of chiral N\'{e}el walls stabilized by interfacial Dzyaloshinskii-Moriya interaction (DMI) whose strength increases going from Pt to Al top layers. This is explained by the presence of DMI with opposite sign at the Pt/Co and Co/M interfaces, the latter increasing in strength going towards heavier atoms, possibly due to the increasing spin-orbit interaction. This work shows that in non-centrosymmetric trilayers the domain wall dynamics can be finely tuned by engineering the DMI strength, in view of efficient devices for logic and spitronics applications.
\end{abstract}

\maketitle

Asymmetric magnetic stacks in which a thin magnetic layer is sandwiched between two heavy metals or a heavy metal and an oxide, hold promise for applications in the field of spintronics or spinorbitronics. They can host chiral domain walls \cite{Thiaville2012}  and skyrmions \cite{Skyrme1960} that can be displaced efficiently with fields or current pulses \cite{Ryu2013,Emori2013,Fert2017}, so that they are envisaged as carriers of binary information in logic devices and racetrack memories \cite{Parkin2008}. The essential ingredient necessary to stabilize such objects  is a strong interfacial Dzyaloshinskii-Moriya interaction \cite{Dzyaloshinskii1957,Moriya1960}, an antisymmetric exchange energy term in competition with Heisenberg exchange that can lead to spiral magnetic structures with a defined chirality.

In order to optimize materials for efficient transfer of digital information, the DMI strength should be controlled and possibly tuned. Nowadays, little is known about the microscopic origins of such interfacial interaction and how it changes as a function of the $3d$ or the $5d$ metal at $3d/5d$ interfaces. Some \textit{ab initio} calculations were performed to address quantitatively the DMI in several HM/FM bilayer systems \cite{Freimuth2014,Yang2015,Belabbes2016} and models have been proposed to explain the DMI variation for several $3d$/$5d$ interfaces \cite{Kashid2014,Belabbes2016}. These calculations can guide the optimization of materials with large DMI, although \textit{ab initio} calculations consider ideal interfaces, while experiments concentrate on samples generally grown by magnetron sputtering, where interfacial structure and quality influences the strength and even the sign of the DMI \cite{Wells2017}. For these reasons, it is rare that experimental findings match the theoretical predictions.

The Pt/Co interface is the prototypical interface hosting a large DMI.  Theoretical \cite{Yang2015,Yang2016,Freimuth2014} and experimental  studies \cite{Pizzini2014,Belmeguenai2015,Cho2015,Kim2016} agree on the fact that it is the source of strong DMI favoring homochiral magnetic textures (chiral N\'{e}el walls and skyrmions) with anticlockwise rotation of the magnetic moments (left-handed chirality). Very little is known about the  DMI at interfaces where Co is in contact with other heavy metals. Only the Pt/Co/Ir stack has been largely studied, following the  \textit{ab initio} calculations by Yang \textit{et al.} that predicted opposite signs for the DMI at Pt(111)/Co and Ir(111)/Co interfaces \cite{Yang2015}. However, experimental data do not allow concluding on this matter. Moreau-Luchaire \textit{et al.} \cite{Moreau-Luchaire2016} reported high DMI values for Pt/Co/Ir multilayers, that they attributed to the same sign of the DMI at Pt/Co and Co/Ir interfaces.  The work of Chen \textit{et al.} \cite{Chen2013} and that of Hrabec  \textit{et al.} \cite{Hrabec2014}  also agree with the theoretical prediction. On the other hand Kim \textit{et al.} \cite{Kim2016} show the same sign of the DMI in Pt/Co/AlOx and Ir/Co/AlOx trilayers, while Han \textit{et al.} \cite{Han2016} found lower DMI for Pt/Co/Ir compared to Pt/Co/AlOx and suggested that this might be due to the opposite DMI signs at Pt/Co and Co/Ir  interfaces. A consensus is still missing on this matter and the contradictory results may be due to the different morphology of the interface between Co and heavy metal in the various materials.

In this work we address the study of DMI in epitaxial layers with well defined (111) texture, like that considered in the \textit{ab initio} work of Ref.\cite{Yang2015}.  The DMI of Pt/Co/M trilayers with different metallic overlayers (M=Al, Cu, Ir, Pt) was studied through the measurement of domain wall dynamics.  We show that the DMI at the Co/M interface has the same sign for all the studied overlayers, with increasing strength as the atomic number increases. The domain wall speed in Co is strongly dependent on the DMI strength, the largest speed being obtained for Pt/Co/Al, the smallest for Pt/Co/Pt where the DMI at the two Co interfaces compensate.
This finding may be useful for engineering samples with optimized DMI for spintronic applications.

The  Pt/Co/M (with M=Al,~Cu,~Ir,~Pt) epitaxial stacks were grown on commercial MgO(111) single crystals.  The morphology of the (111) surface was improved by annealing at high temperature. After one hour of heat treatment at 450$^{\circ}$C, 28 nm of Pt were deposited by dc magnetron sputtering using 6x10$^{-3}$ mbar Ar$^{+}$ pressure and 20W magnetron power. The layer thickness was calibrated by a quartz balance  installed in the sputtering chamber. The surface quality was checked \textit{in-situ} by LEED. The LEED patterns acquired for  MgO(111)/Pt indicate good surface quality with hexagonal reconstruction (Figure 1). The XRD scans covering the region of the FCC Pt(111) crystallographic peak clearly shows that only a single-phase FCC with [111] orientation is present, with no other structural domains coexisting. The $\sim$0.6nm-thick Co, the 2nm-thick M layers and the 2nm-thick Pt capping were grown in general at RT, in order to avoid intermixing at the interfaces. The (111) texture is kept after the Co deposition (Figure 1). For some of the samples, the Co and M layers were grown at 100$^{\circ}$C. We also compared Pt(28nm)/Ir(2nm)/Co(0.6nm)/Pt(2nm) and Pt(30nm)/Co(0.6)/Ir(2nm)/Pt(2nm) samples to check the influence of inversion of the structure on the DMI. The magnetic properties were measured with VSM-SQUID and magneto-optical Kerr effect.  All the samples present perpendicular magnetic anisotropy (PMA) with square hysteresis loops. The unit surface magnetization $M_{s}t$ ($t$ being the thickness of the Co layer) and the in-plane saturation field $\mu_{0}H_{K}$ are reported in Table 1.

\begin{figure}
    \includegraphics[width=8cm]{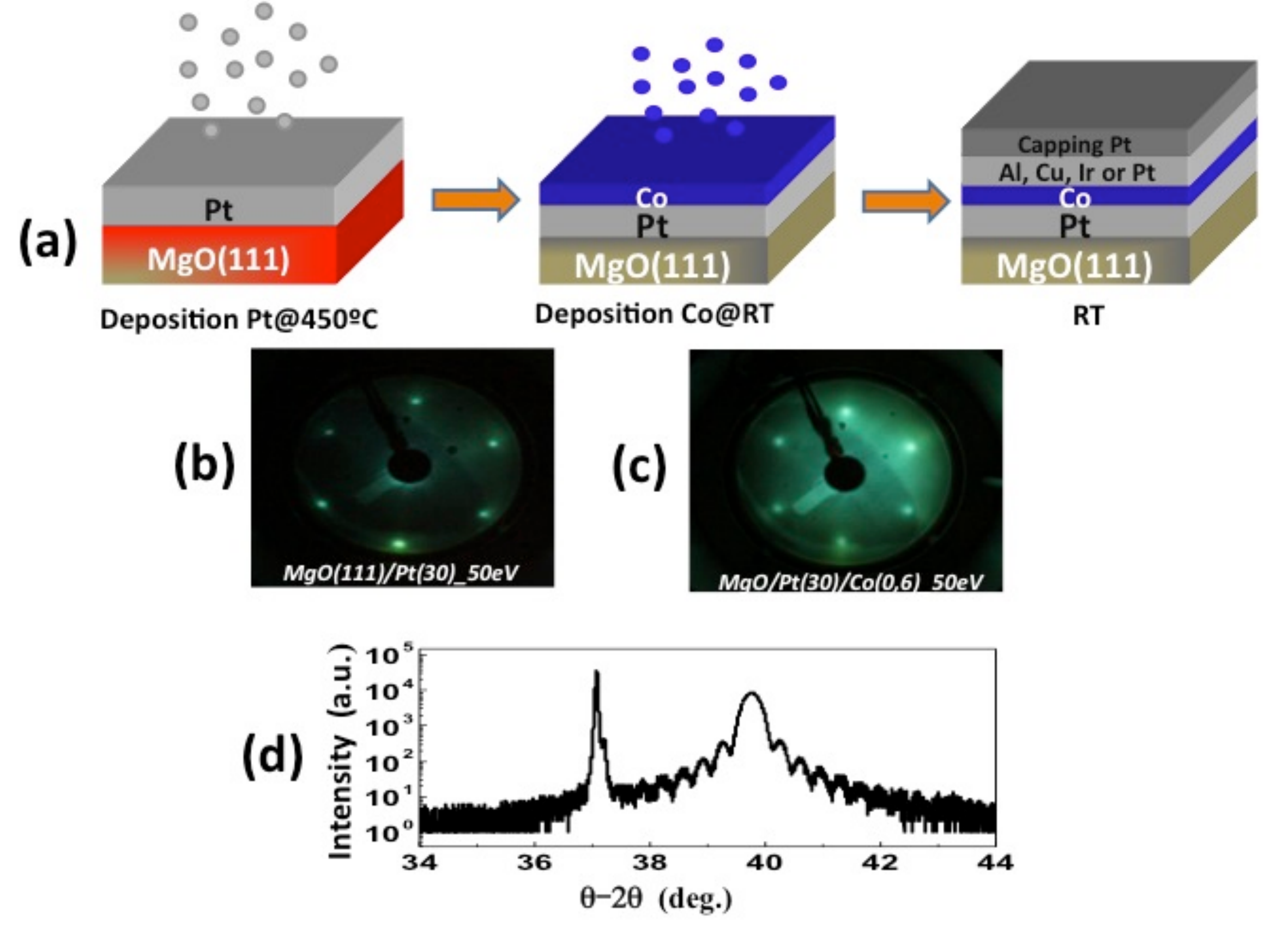}
\caption{\label{fig:Figure1}(a) Sketch of the sample growth process; (b) LEED patterns obtained after the growth of the Pt underlayer; (c) LEED pattern after the growth of the Co layer; (d) XRD pattern of the Pt/Co/Pt trilayer.}
\end{figure}

The domain wall dynamics was studied using wide field magneto-optical Kerr microscopy. The DW velocity was deduced from the expansion of bubble domains, driven by out-of-plane magnetic field pulses $H_{z}$ of strength up to 600 mT, obtained with microcoils associated to pulsed current generators.

The sign and the strength of the DMI for the Pt/Co/M samples was extracted by studying the domain wall expansion driven by $H_{z}$ pulses, in the presence of a constant in-plane magnetic field $H_{x}$ parallel to the DW normal \cite{Je2013,Hrabec2014,Vanatka2015,Pham2016}. In systems with DMI and chiral N\'{e}el walls, the DW propagation is anisotropic in the direction of $H_{x}$ and the DW speed is larger for the domain walls having magnetization parallel to $H_{x}$. This allows us to obtain without ambiguity the chirality of the domain walls.
The DMI strength can be obtained  from the $H_{z}$-driven  DW speed \textit{vs.} the intensity of $H_{x}$.  In the presence of DMI, the DW speed reaches a minimum  when $H_{x}$ compensates the $H_{DMI}$ field that stabilizes the N\'{e}el walls. From the value of this field we can then deduce the average DMI energy density $D$, since
  $H_{DMI} = D /(\mu_{0}M_{s} \Delta)$
where $\Delta = \sqrt{A/K_{0}}$, $A$ is the exchange stiffness and $K_{0}$ the effective anisotropy energy. The $H_{z}$ field driving the DWs was chosen to be beyond the depinning field, giving rise to a reliable measurement of the DMI \cite{Vanatka2015,Pham2016}.

\begin{figure}
    \includegraphics[width=6.5cm]{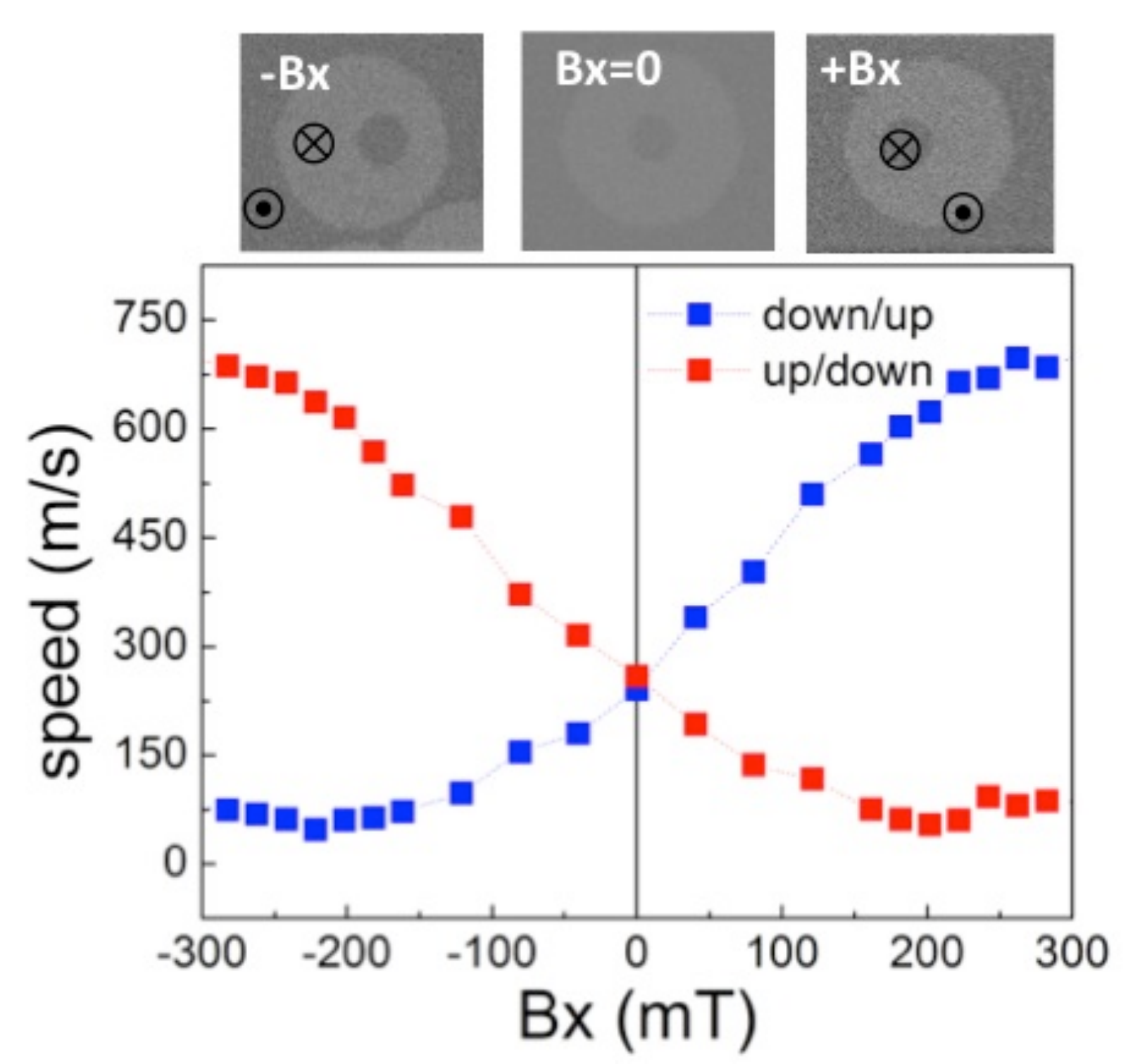}
\caption{\label{fig:Figure2} (Top): Differential Kerr images showing the asymmetric propagation of up/down and down/up DWs in the direction of the $B_{x}$ field in Pt/Co/Al, compared with the isotropic propagation when $B_{x}$=0.  (Bottom): Domain wall speed \textit{vs.} $B_{x}$ for up/down and down/up DWs propagating in the x-direction. The driving field is $B_{z}$=300mT.}
\end{figure}

\begin{figure}
    \includegraphics[width=8.5cm]{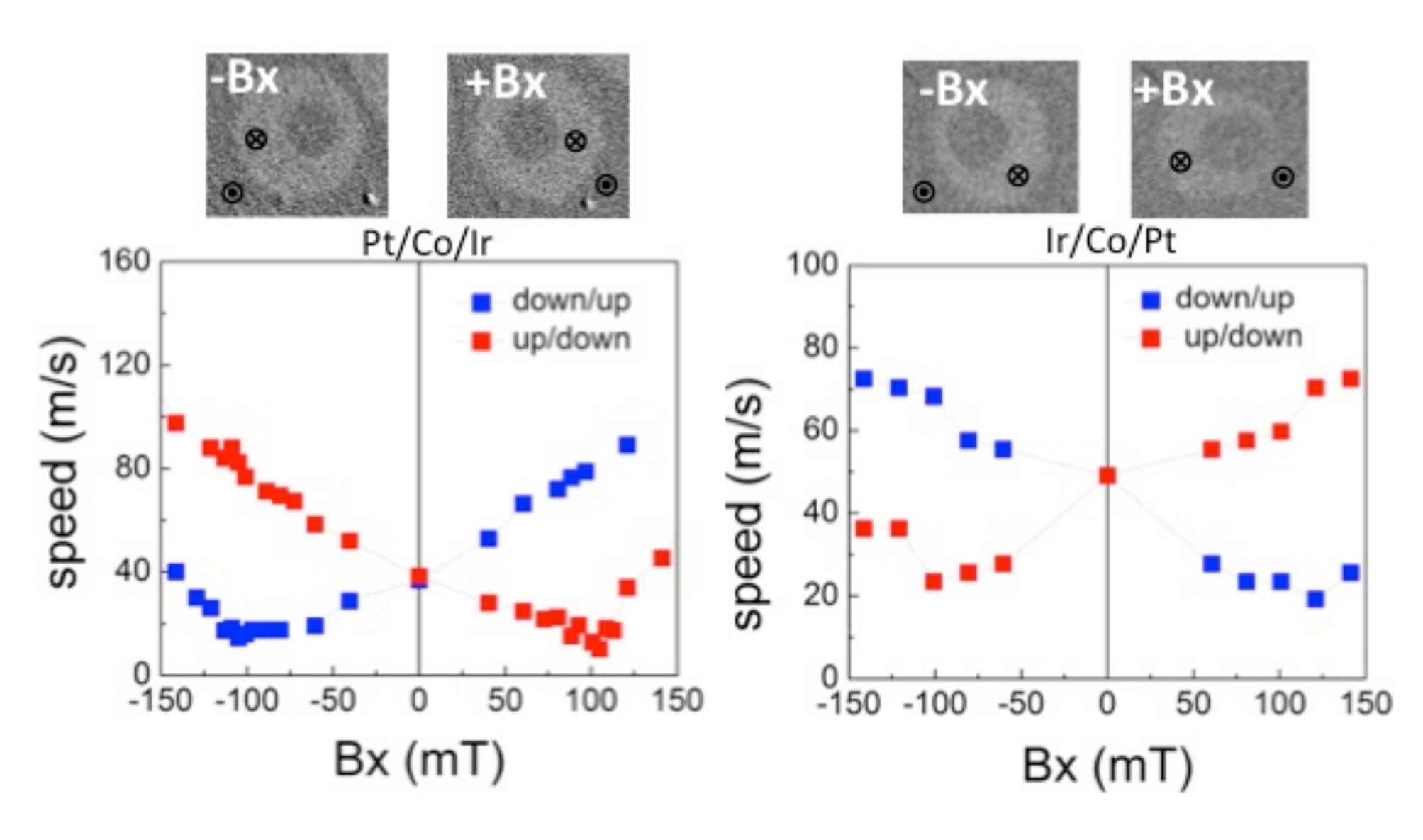}
\caption{\label{fig:Figure3} (Top) Differential Kerr images showing the asymmetric propagation of domain walls in Pt/Co/Ir (left) and Ir/Co/Pt (right), with a driving field $B_{z}$ around 290mT and an in-plane field $B_{x}$ around 100mT.  (Bottom): speed \textit{vs.} $B_{x}$ for up/down and down/up DWs propagating in the x-direction. Note the opposite behaviour of up/down and down/up DWs in the two samples, demonstrating the opposite chirality of the DWs.}
\end{figure}

Figures \ref{fig:Figure2} and \ref{fig:Figure3} present the differential Kerr images showing the anisotropic expansion of bubble domains driven by an $H_{z}$ pulse in the presence of an $H_{x}$ field, measured for Pt/Co/Al, Pt/Co/Ir and Ir/Co/Pt stacks. The same behaviour was found for Pt/Co/Cu trilayers, as expected for systems with chiral N\'{e}el walls. On the other hand, in Pt/Co/Pt stack the DW expansion is perfectly isotropic in the sample plane, as also found for polycrystalline Pt/Co/Pt samples \cite{Pham2016}. This is consistent with the absence of inversion asymmetry, that leads to negligible DMI and to the stabilization of achiral Bloch walls.

Note that the down/up domain walls propagate with larger velocity for positive $H_{x}$  fields  for Pt/Co/M with M=Al,~Cu,~Ir but with smaller velocity for Ir/Co/Pt. These measurements are consistent with the presence of left handed chirality for the  domain walls in Pt/Co/Al, Pt/Co/Cu and Pt/Co/Ir, and an opposite, right-handed, chirality for DWs in Ir/Co/Pt. This is expected using symmetry arguments, as the layer sequence is inverted with respect to the growth axis along which the symmetry is broken.

\begin{table*}
\caption{ Effective anisotropy field  $\mu_{0}H_{K}$, unit surface magnetization $M_{s}t$, saturation domain wall speed $v_{max}$, DMI field $\mu_{0}H_{DMI}$, effective interface DMI energy density extracted from $v_{max}$ (D$_{s}$$^{eff,v}$) and from the DMI field (D$_{s}$$^{eff,H}$), DMI energy density at the Pt/Co interface (D$_{s}$$^{Pt/Co}$) and at the Co/M interface (D$_{s}$$^{Co/M}$) for the Pt/Co/M stacks. Positive DMI values in the table stand for left-handed DW chiralities. }
    \begin{tabular}{p{2cm}  p{1.5cm} p{1.5cm}  p{1.5cm}  p{1.2cm}  p{1.5cm} p{1.5cm}  p{1.5cm} p{1.5cm}  p{1.5cm}   }

    Sample & T$_{Co}$ & $\mu_{0}$H$_{K}$ & M$_{s}$t  & v$_{max}$ & $\mu_{0}$H$_{DMI}$ & D$_{s}$$^{eff,v}$ & D$_{s}$$^{eff,H}$ & D$_{s}$$^{Pt/Co}$ & D$_{s}$$^{Co/M}$ \\
     & ($^{\circ}$C) &(T) & (mA)   & (m/s) & (mT) & (pJ/m)  & (pJ/m) & (pJ/m) & (pJ/m) \\

    \hline

    Pt/Co/Al & RT  & 1.3$\pm$0.1 & 1.3$\pm$0.1   & 280$\pm$10 & 212$\pm$10 & 1.32$\pm$0.1 & 1.21$\pm$0.1 & 1.26$\pm$0.1 & 0 \\

   Pt/Co/Al & 100 & 1.8 & 0.92  & 250 & 200 & 0.83 & 0.91 & 0.87 & 0 \\

    Pt/Co/Ir & RT  & 1.4 & 0.75  & 80 & 96 & 0.22 & 0.30 & 1.26 & -1.0 \\

    Pt/Co/Ir & 100  & 1.3 & 1.0   & 115 & 125 & 0.40 & 0.55  & 0.87 & -0.39\\

    Ir/Co/Pt & RT  & 0.6 & 0.9 & 200 & 106 & -0.65 & -0.65  \\

     Pt/Co/Cu & 100  & 1.8 & 0.75   & 230 & 200 & 0.62 & 0.56 & 0.87 & -0.27\\

     Pt/Co/Pt & 100 & 1.2 & 1.3   & 20 & 0.00 & - & - & \textsl{0.87} & \textsl{-0.87} \\

    \hline

    \end{tabular}

\end{table*}

Figures \ref{fig:Figure2} and \ref{fig:Figure3} also show that the speed of up/down and down/up DWs measured \textit{vs.} the $H_{x}$ intensity reaches a minimum, for a field corresponding to the in-plane field compensating $H_{DMI}$. The strength of the interfacial DMI strength $D_{s}$=$Dt$ (in pJ/m) was obtained from the expression of the DMI field given above, using the measured magnetic parameters and exchange stiffness A=16 pJ/m.

The DMI strengths obtained for the various samples are shown in Table 1 together with the measured magnetic parameters. Data are shown for samples with Co grown both at RT and at 100$^{\circ}$C. Let us first look at the DMI values for Co grown at RT. The largest value of the DMI is obtained for Pt/Co/Al ($D_{s}$$^{eff}$=1.2-1.3~pJ/m), while it falls to $D_{s}$$^{eff}$=0.2-0.3~pJ/m for Pt/Co/Ir and is vanishing for Pt/Co/Pt. If we consider that the two Co interfaces are sufficiently apart so that they give a distinct contribution to the DMI \cite{Yang2015}, then $D_{s}$$^{eff}$ = $D_{s}$$^{Pt/Co/M}$ =$D_{s}$$^{Pt/Co}$+$D_{s}$$^{Co/M}$.  Since for all the samples the growth process was identical up to the Co layer, we can assume that the DMI strength at the Pt/Co interface does not change from one sample to another. These results then indicate that the DMI decrease going from Al to Ir and to Pt, is due to the consequent increase of the contribution of the Co/M interface, its sign being opposite to that of the  Pt/Co interface. Since Al has a low atomic number and therefore a weak SOC, we suppose that the DMI at the Co/Al interface is negligible and that the DMI of the Pt/Co/Al trilayer is concentrated at the Pt/Co interface. This is corroborated by our recent measurements of a series of Pt/Co/AlOx samples with variable oxygen concentration at the top Co interface \cite{Chaves2017}. While the under-oxidized top interface, where Co is essentially in contact with Al, does not contribute to the DMI and to the PMA, a gradual increase of the DMI occurs as the Al atoms are substituted by O. The value of the interfacial DMI for the under-oxidized Pt/Co/AlOx sample was found to be around 1.2~pJ/m in agreement with the results reported here.

Using these arguments, our results suggest that the  small value of the DMI in the Pt/Co/Ir trilayer is due to a large DMI contribution of the Co/Ir interface ($D_{s}$$^{Co/Ir}$~=~-1~pJ/m),  opposite to that of the Pt/Co interface $D_{s}$$^{Pt/Co}$~=~1.2-1.3~pJ/m). This is in agreement with the results of Kim \textit{et al.} \cite{Kim2016} who find the same DMI sign at Pt/Co and Ir/Co interfaces in polycrystalline HM/Co/AlOx (HM=Pt,~Ir) trilayers.  In their case the DMI at the Ir/Co interface is a factor 3 smaller than that of the Pt/Co interface: this may be attributed to the different details of the HM/Co interface morphology, which can strongly influence the DMI. The better crystallinity of our samples may explain the better "compensation"  of the DMI at the two interfaces (larger DMI at Co/Ir), which is realized perfectly in the case of Pt/Co/Pt.

Note that for Pt/Co/Al, the DMI decreases to $D_{s}$$^{eff}$$\sim$~0.9~pJ/m when the Co layer is grown at 100$^{\circ}$C. Since the Co/Al interface does not give rise to DMI, this deterioration of the DMI is probably due to the slight interdiffusion  at the  Pt/Co interface. A change occurs also for the DMI of Pt/Co/Ir grown at high temperature, but in this case the effective DMI increases, as the different modification of the DMI at the two interfaces gives rise to a worse compensation. Assuming again that $D_{s}^{Pt/Co}$$\sim$~0.9~pJ/m also in this trilayer, then the DMI at the Co/Ir is shown to decrease to $D_{s}$$^{Co/Ir}$~$\sim$~-0.4~pJ/m when Co is grown at high temperature. These results show that the morphology of the interface is as expected  an important parameter that can strongly influence the strength of the DMI (see also \cite{Wells2017}).

The Pt/Co/Cu sample has an intermediate behavior between Pt/Co/Al and Pt/Co/Ir. For technical reasons only the sample with Co grown at 100$^{\circ}$C could be measured, for which we found $D_{s}$$^{eff}$$\sim$~0.6~pJ/m.  Using the same assumption for the Pt/Co interface, we deduce that $D_{s}^{Co/Cu}$$\sim$-0.3~pJ/m, that is smaller than the value obtained for the Co/Ir interface grown at the same temperature. Since Co and Cu are immiscible at 100$^{\circ}$C, we do not expect a large variation of this value for a sample with Co grown at RT.

The observed increase of the DMI at the Co/M interface, with M going from Al to Cu to Ir and finally to Pt may be related to the increasing atomic number and therefore to the increasing SOC, which is known to be at the origin of the DMI.

These results show that the DMI of these stacks can be easily manipulated and controlled by varying the chemical nature of the top layer. This can provide an interesting step towards the realization of \textit{ad hoc} materials for spintronic devices, in which the DMI is an essential parameter \textit{e.g.} to stabilize magnetic skyrmions.

\begin{figure}
    \includegraphics[width=6cm]{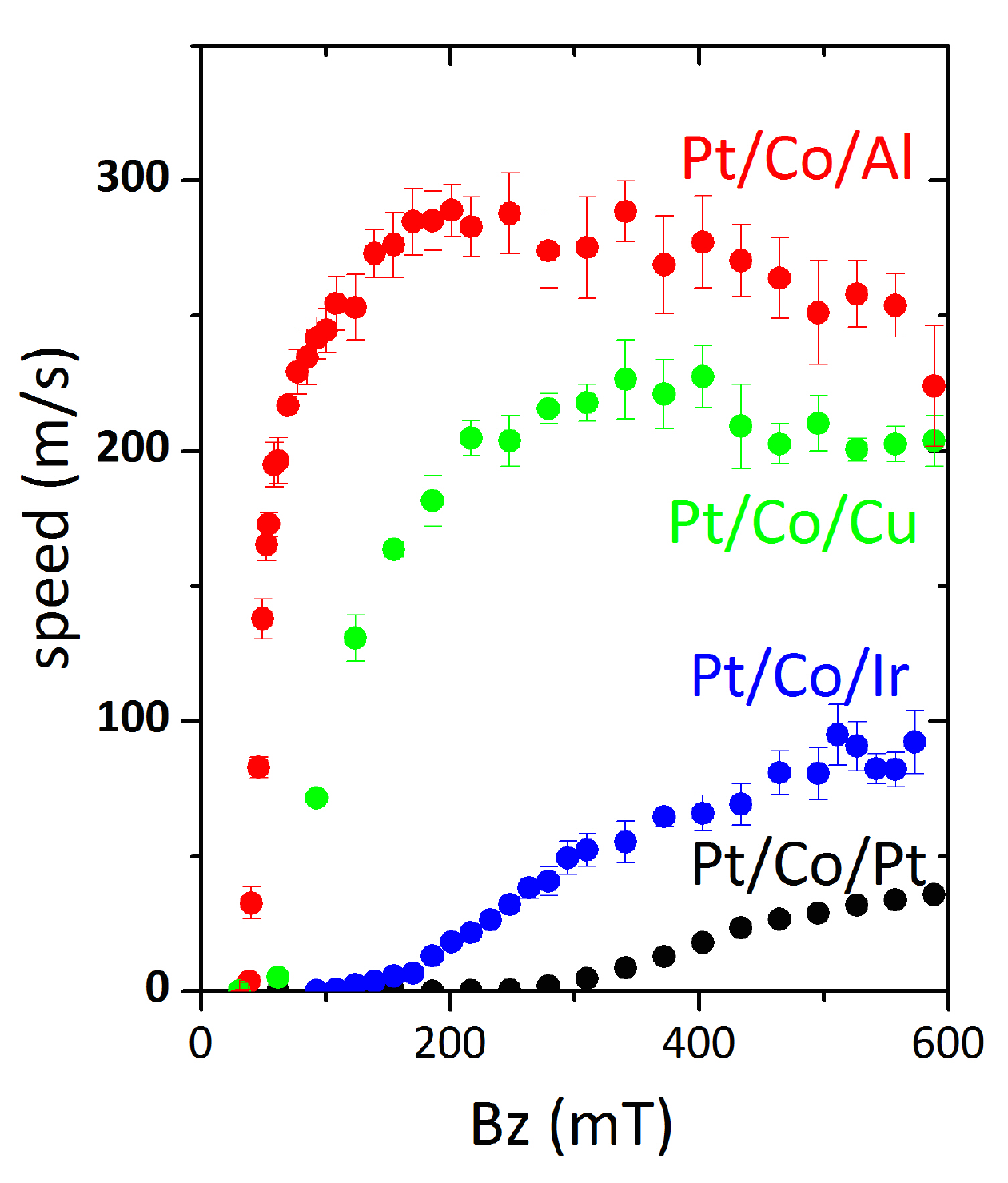}
\caption{\label{fig:Figure4}(a) Domain wall speed vs. out-of-plane magnetic field, measured for Pt/Co/Pt, Pt/Co/Ir, Pt/Co/Cu and Pt/Co/Al. Note the increasing saturation speed as the DMI gets larger.}
\end{figure}

We also investigated the effect of the effective DMI variation on the domain wall velocity \textit{vs.} out-of-plane field. This is shown in Figure 4. The DW dynamics  strongly depends on the chemical nature of the top layer, the  largest velocity  obtained for Pt/Co/Al (maximum DMI)  being a factor 10 larger than  for Pt/Co/Pt (vanishing DMI) in the same field range.  This result is related to the fact that in samples with DMI the saturation DW velocity is linearly proportional to $D_{S}$, as $v_{max} = \frac{\pi}{2} \gamma \frac{D_{s}}{M_{s}t}$ \cite{Pham2016,Chaves2017}. Table 1 shows the values of the interfacial DMI found using this analytical expression and the experimental $v_{max}$ and unit surface magnetization $M_{s}t$. These values are in excellent agreement with those found with the method used above.

In conclusion, we have shown that the DMI and, as a consequence, the domain wall velocity in Pt/Co/M trilayers can be tuned by changing the chemical species of the top metallic layer. This provides an interesting way to tune the DMI of trilayer system in view of optimized spintronics or logic devices.

In these epitaxial samples with strong (111) texture, the DMI at the Co/M interface (with M=Al,~Cu,~Ir and~Pt) was found to have opposite sign to that of the Pt/Co interface, and to increase in strength for heavier atoms, probably because of the increasing SOC. Since the effective DMI is the sum of the DMI at top and bottom Co interfaces, its maximum strength (and maximum DW speed) is obtained for Pt/Co/Al trilayers where the Co/Al interface gives a negligible contribution to the DMI. The microscopic reasons for the disagreement between these measurements and the results of \textit{ab initio}  calculations are still to be revealed, but our results point out the importance of the details of the interface structure and its effects on the electronic hybridization.

Preliminary data on Co/Ta and Co/W layers grown on Pt(111) suggest a  DMI lower and of opposite sign of that of the Pt/Co interface, like for Co/Al, Co/Cu and Co/Ir interfaces. These data suggest that among the interfaces investigated to date, the Pt/Co is the one providing the largest DMI. An efficient way to increase the DMI of a complete trilayer, is to cover the Co layer with an optimized concentration of oxygen \cite{Chaves2017}.

\begin{acknowledgments}S.~P. and J.~V. acknowledge the support of the Agence Nationale de la Recherche, project ANR-14-CE26-0012 (ULTRASKY). B. Fernandez, Ph. David and E. Mossang are acknowledged for their technical help. D.S.C was supported by a CNPq Scholarship (Brazil). F. A. acknowledges the support of Spanish MINECO for exchange grant EEBB-I-16-11844. P.P., R.G., A.G. and J.C. acknowledge the support of Spanish MINECO project FIS2016-78591-C3-1-R (SKYTRON). F.A., P.P. and J.C. acknowledge the support of the Comunidad de Madrid through project NANOFRONTMAG CM.
\end{acknowledgments}


\begin{thebibliography}{26}%
\makeatletter
\providecommand \@ifxundefined [1]{%
 \@ifx{#1\undefined}
}%
\providecommand \@ifnum [1]{%
 \ifnum #1\expandafter \@firstoftwo
 \else \expandafter \@secondoftwo
 \fi
}%
\providecommand \@ifx [1]{%
 \ifx #1\expandafter \@firstoftwo
 \else \expandafter \@secondoftwo
 \fi
}%
\providecommand \natexlab [1]{#1}%
\providecommand \enquote  [1]{``#1''}%
\providecommand \bibnamefont  [1]{#1}%
\providecommand \bibfnamefont [1]{#1}%
\providecommand \citenamefont [1]{#1}%
\providecommand \href@noop [0]{\@secondoftwo}%
\providecommand \href [0]{\begingroup \@sanitize@url \@href}%
\providecommand \@href[1]{\@@startlink{#1}\@@href}%
\providecommand \@@href[1]{\endgroup#1\@@endlink}%
\providecommand \@sanitize@url [0]{\catcode `\\12\catcode `\$12\catcode
  `\&12\catcode `\#12\catcode `\^12\catcode `\_12\catcode `\%12\relax}%
\providecommand \@@startlink[1]{}%
\providecommand \@@endlink[0]{}%
\providecommand \url  [0]{\begingroup\@sanitize@url \@url }%
\providecommand \@url [1]{\endgroup\@href {#1}{\urlprefix }}%
\providecommand \urlprefix  [0]{URL }%
\providecommand \Eprint [0]{\href }%
\providecommand \doibase [0]{http://dx.doi.org/}%
\providecommand \selectlanguage [0]{\@gobble}%
\providecommand \bibinfo  [0]{\@secondoftwo}%
\providecommand \bibfield  [0]{\@secondoftwo}%
\providecommand \translation [1]{[#1]}%
\providecommand \BibitemOpen [0]{}%
\providecommand \bibitemStop [0]{}%
\providecommand \bibitemNoStop [0]{.\EOS\space}%
\providecommand \EOS [0]{\spacefactor3000\relax}%
\providecommand \BibitemShut  [1]{\csname bibitem#1\endcsname}%
\let\auto@bib@innerbib\@empty
\bibitem [{\citenamefont {Thiaville}\ \emph {et~al.}(2012)\citenamefont
  {Thiaville}, \citenamefont {Rohart}, \citenamefont {Ju{\'{e}}}, \citenamefont
  {Cros},\ and\ \citenamefont {Fert}}]{Thiaville2012}%
  \BibitemOpen
  \bibfield  {author} {\bibinfo {author} {\bibfnamefont {A.}~\bibnamefont
  {Thiaville}}, \bibinfo {author} {\bibfnamefont {S.}~\bibnamefont {Rohart}},
  \bibinfo {author} {\bibfnamefont {E.}~\bibnamefont {Ju{\'{e}}}}, \bibinfo
  {author} {\bibfnamefont {V.}~\bibnamefont {Cros}}, \ and\ \bibinfo {author}
  {\bibfnamefont {A.}~\bibnamefont {Fert}},\ }\href {\doibase
  10.1209/0295-5075/100/57002} {\bibfield  {journal} {\bibinfo  {journal}
  {EPL}\ }\textbf {\bibinfo {volume} {100}},\ \bibinfo {pages} {57002}
  (\bibinfo {year} {2012})}\BibitemShut {NoStop}%
\bibitem [{\citenamefont {Skyrme}(1962)}]{Skyrme1960}%
  \BibitemOpen
  \bibfield  {author} {\bibinfo {author} {\bibfnamefont {T.}~\bibnamefont
  {Skyrme}},\ }\href@noop {} {\bibfield  {journal} {\bibinfo  {journal} {Nucl.
  Phys.}\ }\textbf {\bibinfo {volume} {31}},\ \bibinfo {pages} {556} (\bibinfo
  {year} {1962})}\BibitemShut {NoStop}%
\bibitem [{\citenamefont {Ryu}\ \emph {et~al.}(2013)\citenamefont {Ryu},
  \citenamefont {L.}, \citenamefont {Yang},\ and\ \citenamefont
  {Parkin}}]{Ryu2013}%
  \BibitemOpen
  \bibfield  {author} {\bibinfo {author} {\bibfnamefont {K.-S.}\ \bibnamefont
  {Ryu}}, \bibinfo {author} {\bibfnamefont {T.}~\bibnamefont {L.}}, \bibinfo
  {author} {\bibfnamefont {S.-H.}\ \bibnamefont {Yang}}, \ and\ \bibinfo
  {author} {\bibfnamefont {S.}~\bibnamefont {Parkin}},\ }\href {\doibase
  10.1038/NNANO.2013.102} {\bibfield  {journal} {\bibinfo  {journal} {Nat.
  Nanotech.}\ }\textbf {\bibinfo {volume} {8}},\ \bibinfo {pages} {527}
  (\bibinfo {year} {2013})}\BibitemShut {NoStop}%
\bibitem [{\citenamefont {Emori}\ \emph {et~al.}(2013)\citenamefont {Emori},
  \citenamefont {Bauer}, \citenamefont {Ahn}, \citenamefont {Martinez},\ and\
  \citenamefont {Beach}}]{Emori2013}%
  \BibitemOpen
  \bibfield  {author} {\bibinfo {author} {\bibfnamefont {S.}~\bibnamefont
  {Emori}}, \bibinfo {author} {\bibfnamefont {U.}~\bibnamefont {Bauer}},
  \bibinfo {author} {\bibfnamefont {S.-M.}\ \bibnamefont {Ahn}}, \bibinfo
  {author} {\bibfnamefont {E.}~\bibnamefont {Martinez}}, \ and\ \bibinfo
  {author} {\bibfnamefont {G.}~\bibnamefont {Beach}},\ }\href {\doibase
  {10.1038/NMAT3675}} {\bibfield  {journal} {\bibinfo  {journal} {Nat. Mater.}\
  }\textbf {\bibinfo {volume} {12}},\ \bibinfo {pages} {611} (\bibinfo {year}
  {2013})}\BibitemShut {NoStop}%
\bibitem [{\citenamefont {Fert}, \citenamefont {Cros},\ and\ \citenamefont
  {Reyren}(2017)}]{Fert2017}%
  \BibitemOpen
  \bibfield  {author} {\bibinfo {author} {\bibfnamefont {A.}~\bibnamefont
  {Fert}}, \bibinfo {author} {\bibfnamefont {V.}~\bibnamefont {Cros}}, \ and\
  \bibinfo {author} {\bibfnamefont {N.}~\bibnamefont {Reyren}},\ }\href@noop {}
  {\bibfield  {journal} {\bibinfo  {journal} {Nature Rev. Mater.}\ }\textbf
  {\bibinfo {volume} {2}},\ \bibinfo {pages} {17031} (\bibinfo {year}
  {2017})}\BibitemShut {NoStop}%
\bibitem [{\citenamefont {Parkin}, \citenamefont {Hayashi},\ and\ \citenamefont
  {Thomas}(2008)}]{Parkin2008}%
  \BibitemOpen
  \bibfield  {author} {\bibinfo {author} {\bibfnamefont {S.}~\bibnamefont
  {Parkin}}, \bibinfo {author} {\bibfnamefont {M.}~\bibnamefont {Hayashi}}, \
  and\ \bibinfo {author} {\bibfnamefont {L.}~\bibnamefont {Thomas}},\ }\href
  {\doibase {10.1126/science.1145799}} {\bibfield  {journal} {\bibinfo
  {journal} {Science}\ }\textbf {\bibinfo {volume} {320}},\ \bibinfo {pages}
  {190} (\bibinfo {year} {2008})}\BibitemShut {NoStop}%
\bibitem [{\citenamefont {Dzyaloshinskii}(1957)}]{Dzyaloshinskii1957}%
  \BibitemOpen
  \bibfield  {author} {\bibinfo {author} {\bibfnamefont {I.~E.}\ \bibnamefont
  {Dzyaloshinskii}},\ }\href@noop {} {\bibfield  {journal} {\bibinfo  {journal}
  {Sov. Phys. JETP}\ }\textbf {\bibinfo {volume} {5}},\ \bibinfo {pages} {1259}
  (\bibinfo {year} {1957})}\BibitemShut {NoStop}%
\bibitem [{\citenamefont {Moriya}(1960)}]{Moriya1960}%
  \BibitemOpen
  \bibfield  {author} {\bibinfo {author} {\bibfnamefont {T.}~\bibnamefont
  {Moriya}},\ }\href@noop {} {\bibfield  {journal} {\bibinfo  {journal} {Phys.
  Rev.}\ }\textbf {\bibinfo {volume} {120}},\ \bibinfo {pages} {91} (\bibinfo
  {year} {1960})}\BibitemShut {NoStop}%
\bibitem [{\citenamefont {Freimuth}, \citenamefont {Bl\"ugel},\ and\
  \citenamefont {Mokrousov}(2014)}]{Freimuth2014}%
  \BibitemOpen
  \bibfield  {author} {\bibinfo {author} {\bibfnamefont {F.}~\bibnamefont
  {Freimuth}}, \bibinfo {author} {\bibfnamefont {S.}~\bibnamefont {Bl\"ugel}},
  \ and\ \bibinfo {author} {\bibfnamefont {Y.}~\bibnamefont {Mokrousov}},\
  }\href@noop {} {\bibfield  {journal} {\bibinfo  {journal} {J. Phys.: Condens.
  Matter}\ }\textbf {\bibinfo {volume} {26}},\ \bibinfo {pages} {104202}
  (\bibinfo {year} {2014})}\BibitemShut {NoStop}%
\bibitem [{\citenamefont {Yang}\ \emph {et~al.}(2015)\citenamefont {Yang},
  \citenamefont {Thiaville}, \citenamefont {Rohart}, \citenamefont {Fert},\
  and\ \citenamefont {Chshiev}}]{Yang2015}%
  \BibitemOpen
  \bibfield  {author} {\bibinfo {author} {\bibfnamefont {H.}~\bibnamefont
  {Yang}}, \bibinfo {author} {\bibfnamefont {A.}~\bibnamefont {Thiaville}},
  \bibinfo {author} {\bibfnamefont {S.}~\bibnamefont {Rohart}}, \bibinfo
  {author} {\bibfnamefont {A.}~\bibnamefont {Fert}}, \ and\ \bibinfo {author}
  {\bibfnamefont {M.}~\bibnamefont {Chshiev}},\ }\href {\doibase
  10.1103/PhysRevLett.115.267210} {\bibfield  {journal} {\bibinfo  {journal}
  {Phys. Rev. Lett.}\ }\textbf {\bibinfo {volume} {115}},\ \bibinfo {pages}
  {267210} (\bibinfo {year} {2015})}\BibitemShut {NoStop}%
\bibitem [{\citenamefont {Belabbes}, \citenamefont {Bl\"ugel},\ and\
  \citenamefont {Manchon}(2016)}]{Belabbes2016}%
  \BibitemOpen
  \bibfield  {author} {\bibinfo {author} {\bibfnamefont {A.}~\bibnamefont
  {Belabbes}}, \bibinfo {author} {\bibfnamefont {S.}~\bibnamefont {Bl\"ugel}},
  \ and\ \bibinfo {author} {\bibfnamefont {A.}~\bibnamefont {Manchon}},\
  }\href@noop {} {\bibfield  {journal} {\bibinfo  {journal} {Phys. Rev. Lett.}\
  }\textbf {\bibinfo {volume} {117}},\ \bibinfo {pages} {247202} (\bibinfo
  {year} {2016})}\BibitemShut {NoStop}%
\bibitem [{\citenamefont {Kashid}\ \emph {et~al.}(2014)\citenamefont {Kashid},
  \citenamefont {Schena}, \citenamefont {Zimmermann}, \citenamefont
  {Mokrousov}, \citenamefont {Bl\"ugel}, \citenamefont {Shah},\ and\
  \citenamefont {Salunke}}]{Kashid2014}%
  \BibitemOpen
  \bibfield  {author} {\bibinfo {author} {\bibfnamefont {V.}~\bibnamefont
  {Kashid}}, \bibinfo {author} {\bibfnamefont {T.}~\bibnamefont {Schena}},
  \bibinfo {author} {\bibfnamefont {B.}~\bibnamefont {Zimmermann}}, \bibinfo
  {author} {\bibfnamefont {Y.}~\bibnamefont {Mokrousov}}, \bibinfo {author}
  {\bibfnamefont {S.}~\bibnamefont {Bl\"ugel}}, \bibinfo {author}
  {\bibfnamefont {V.}~\bibnamefont {Shah}}, \ and\ \bibinfo {author}
  {\bibfnamefont {H.~G.}\ \bibnamefont {Salunke}},\ }\href {\doibase
  10.1103/PhysRevB.90.054412} {\bibfield  {journal} {\bibinfo  {journal} {Phys.
  Rev. B}\ }\textbf {\bibinfo {volume} {90}},\ \bibinfo {pages} {054412}
  (\bibinfo {year} {2014})}\BibitemShut {NoStop}%
\bibitem [{\citenamefont {Wells}\ \emph {et~al.}(2017)\citenamefont {Wells},
  \citenamefont {Shepley}, \citenamefont {Marrows},\ and\ \citenamefont
  {Moore}}]{Wells2017}%
  \BibitemOpen
  \bibfield  {author} {\bibinfo {author} {\bibfnamefont {A.~W.~J.}\
  \bibnamefont {Wells}}, \bibinfo {author} {\bibfnamefont {P.~M.}\ \bibnamefont
  {Shepley}}, \bibinfo {author} {\bibfnamefont {C.~H.}\ \bibnamefont
  {Marrows}}, \ and\ \bibinfo {author} {\bibfnamefont {T.~A.}\ \bibnamefont
  {Moore}},\ }\href {\doibase 10.1103/PhysRevB.95.054428} {\bibfield  {journal}
  {\bibinfo  {journal} {Phys. Rev. B}\ }\textbf {\bibinfo {volume} {95}},\
  \bibinfo {pages} {054428} (\bibinfo {year} {2017})}\BibitemShut {NoStop}%
\bibitem [{\citenamefont {Yang}\ \emph {et~al.}(2016)\citenamefont {Yang},
  \citenamefont {Boulle}, \citenamefont {Cros}, \citenamefont {Fert},\ and\
  \citenamefont {Chshiev}}]{Yang2016}%
  \BibitemOpen
  \bibfield  {author} {\bibinfo {author} {\bibfnamefont {H.}~\bibnamefont
  {Yang}}, \bibinfo {author} {\bibfnamefont {O.}~\bibnamefont {Boulle}},
  \bibinfo {author} {\bibfnamefont {V.}~\bibnamefont {Cros}}, \bibinfo {author}
  {\bibfnamefont {A.}~\bibnamefont {Fert}}, \ and\ \bibinfo {author}
  {\bibfnamefont {M.}~\bibnamefont {Chshiev}},\ }\href@noop {} {\bibfield
  {journal} {\bibinfo  {journal} {arXiv:1603.01847}\ } (\bibinfo {year}
  {2016})}\BibitemShut {NoStop}%
\bibitem [{\citenamefont {Pizzini}\ \emph {et~al.}(2014)\citenamefont
  {Pizzini}, \citenamefont {Vogel}, \citenamefont {Rohart}, \citenamefont
  {Buda-Prejbeanu}, \citenamefont {Ju\'{e}}, \citenamefont {Boulle},
  \citenamefont {Miron}, \citenamefont {Safeer}, \citenamefont {Auffret},
  \citenamefont {Gaudin},\ and\ \citenamefont {Thiaville}}]{Pizzini2014}%
  \BibitemOpen
  \bibfield  {author} {\bibinfo {author} {\bibfnamefont {S.}~\bibnamefont
  {Pizzini}}, \bibinfo {author} {\bibfnamefont {J.}~\bibnamefont {Vogel}},
  \bibinfo {author} {\bibfnamefont {S.}~\bibnamefont {Rohart}}, \bibinfo
  {author} {\bibfnamefont {L.}~\bibnamefont {Buda-Prejbeanu}}, \bibinfo
  {author} {\bibfnamefont {E.}~\bibnamefont {Ju\'{e}}}, \bibinfo {author}
  {\bibfnamefont {O.}~\bibnamefont {Boulle}}, \bibinfo {author} {\bibfnamefont
  {I.}~\bibnamefont {Miron}}, \bibinfo {author} {\bibfnamefont
  {C.}~\bibnamefont {Safeer}}, \bibinfo {author} {\bibfnamefont
  {S.}~\bibnamefont {Auffret}}, \bibinfo {author} {\bibfnamefont
  {G.}~\bibnamefont {Gaudin}}, \ and\ \bibinfo {author} {\bibfnamefont
  {A.}~\bibnamefont {Thiaville}},\ }\href {\doibase
  10.1103/PhysRevLett.113.047203} {\bibfield  {journal} {\bibinfo  {journal}
  {Phys. Rev. Lett.}\ }\textbf {\bibinfo {volume} {113}},\ \bibinfo {pages}
  {047203} (\bibinfo {year} {2014})}\BibitemShut {NoStop}%
\bibitem [{\citenamefont {Belmeguenai}\ \emph {et~al.}(2015)\citenamefont
  {Belmeguenai}, \citenamefont {Adam}, \citenamefont {Roussign\'{e}},
  \citenamefont {Eimer}, \citenamefont {Devolder}, \citenamefont {Kim},
  \citenamefont {Cherif}, \citenamefont {Stashkevich},\ and\ \citenamefont
  {Thiaville}}]{Belmeguenai2015}%
  \BibitemOpen
  \bibfield  {author} {\bibinfo {author} {\bibfnamefont {M.}~\bibnamefont
  {Belmeguenai}}, \bibinfo {author} {\bibfnamefont {J.-P.}\ \bibnamefont
  {Adam}}, \bibinfo {author} {\bibfnamefont {Y.}~\bibnamefont {Roussign\'{e}}},
  \bibinfo {author} {\bibfnamefont {S.}~\bibnamefont {Eimer}}, \bibinfo
  {author} {\bibfnamefont {T.}~\bibnamefont {Devolder}}, \bibinfo {author}
  {\bibfnamefont {J.-V.}\ \bibnamefont {Kim}}, \bibinfo {author} {\bibfnamefont
  {S.}~\bibnamefont {Cherif}}, \bibinfo {author} {\bibfnamefont
  {A.}~\bibnamefont {Stashkevich}}, \ and\ \bibinfo {author} {\bibfnamefont
  {A.}~\bibnamefont {Thiaville}},\ }\href@noop {} {\bibfield  {journal}
  {\bibinfo  {journal} {Phys. Rev. B}\ }\textbf {\bibinfo {volume} {91}},\
  \bibinfo {pages} {180405(R)} (\bibinfo {year} {2015})}\BibitemShut {NoStop}%
\bibitem [{\citenamefont {Cho}\ \emph {et~al.}(2015)\citenamefont {Cho},
  \citenamefont {Kim}, \citenamefont {Lee}, \citenamefont {Kim}, \citenamefont
  {Lavrijsen}, \citenamefont {Solignac}, \citenamefont {Yin}, \citenamefont
  {Han}, \citenamefont {van Hoof}, \citenamefont {Swagten}, \citenamefont
  {Koopmans},\ and\ \citenamefont {You}}]{Cho2015}%
  \BibitemOpen
  \bibfield  {author} {\bibinfo {author} {\bibfnamefont {J.}~\bibnamefont
  {Cho}}, \bibinfo {author} {\bibfnamefont {N.-H.}\ \bibnamefont {Kim}},
  \bibinfo {author} {\bibfnamefont {S.}~\bibnamefont {Lee}}, \bibinfo {author}
  {\bibfnamefont {J.-S.}\ \bibnamefont {Kim}}, \bibinfo {author} {\bibfnamefont
  {R.}~\bibnamefont {Lavrijsen}}, \bibinfo {author} {\bibfnamefont
  {A.}~\bibnamefont {Solignac}}, \bibinfo {author} {\bibfnamefont
  {Y.}~\bibnamefont {Yin}}, \bibinfo {author} {\bibfnamefont {D.-S.}\
  \bibnamefont {Han}}, \bibinfo {author} {\bibfnamefont {N.~J.~J.}\
  \bibnamefont {van Hoof}}, \bibinfo {author} {\bibfnamefont {H.~J.~M.}\
  \bibnamefont {Swagten}}, \bibinfo {author} {\bibfnamefont {B.}~\bibnamefont
  {Koopmans}}, \ and\ \bibinfo {author} {\bibfnamefont {C.-Y.}\ \bibnamefont
  {You}},\ }\href@noop {} {\bibfield  {journal} {\bibinfo  {journal} {{Nature
  Comm.}}\ }\textbf {\bibinfo {volume} {{6}}} (\bibinfo {year}
  {{2015}})}\BibitemShut {NoStop}%
\bibitem [{\citenamefont {Kim}\ \emph {et~al.}(2016)\citenamefont {Kim},
  \citenamefont {Jung}, \citenamefont {Cho}, \citenamefont {Han}, \citenamefont
  {Yin}, \citenamefont {Kim}, \citenamefont {Swagten},\ and\ \citenamefont
  {You}}]{Kim2016}%
  \BibitemOpen
  \bibfield  {author} {\bibinfo {author} {\bibfnamefont {N.-H.}\ \bibnamefont
  {Kim}}, \bibinfo {author} {\bibfnamefont {J.}~\bibnamefont {Jung}}, \bibinfo
  {author} {\bibfnamefont {J.}~\bibnamefont {Cho}}, \bibinfo {author}
  {\bibfnamefont {D.-S.}\ \bibnamefont {Han}}, \bibinfo {author} {\bibfnamefont
  {Y.}~\bibnamefont {Yin}}, \bibinfo {author} {\bibfnamefont {J.-S.}\
  \bibnamefont {Kim}}, \bibinfo {author} {\bibfnamefont {H.~J.~M.}\
  \bibnamefont {Swagten}}, \ and\ \bibinfo {author} {\bibfnamefont {C.-Y.}\
  \bibnamefont {You}},\ }\href {\doibase 10.1063/1.4945685} {\bibfield
  {journal} {\bibinfo  {journal} {Applied Physics Letters}\ }\textbf {\bibinfo
  {volume} {108}},\ \bibinfo {pages} {142406} (\bibinfo {year} {2016})},\
  \Eprint {http://arxiv.org/abs/http://dx.doi.org/10.1063/1.4945685}
  {http://dx.doi.org/10.1063/1.4945685} \BibitemShut {NoStop}%
\bibitem [{\citenamefont {Moreau-Luchaire}\ \emph {et~al.}(2016)\citenamefont
  {Moreau-Luchaire}, \citenamefont {Moutafis}, \citenamefont {Reyren},
  \citenamefont {Sampaio}, \citenamefont {Vaz}, \citenamefont {Van~Horne},
  \citenamefont {Bouzehouane}, \citenamefont {Garcia}, \citenamefont
  {Deranlot}, \citenamefont {Warnicke}, \citenamefont {Wohlhuter},
  \citenamefont {George}, \citenamefont {Weigand}, \citenamefont {Raabe},
  \citenamefont {Cros},\ and\ \citenamefont {Fert}}]{Moreau-Luchaire2016}%
  \BibitemOpen
  \bibfield  {author} {\bibinfo {author} {\bibfnamefont {C.}~\bibnamefont
  {Moreau-Luchaire}}, \bibinfo {author} {\bibfnamefont {C.}~\bibnamefont
  {Moutafis}}, \bibinfo {author} {\bibfnamefont {N.}~\bibnamefont {Reyren}},
  \bibinfo {author} {\bibfnamefont {J.}~\bibnamefont {Sampaio}}, \bibinfo
  {author} {\bibfnamefont {C.~A.~F.}\ \bibnamefont {Vaz}}, \bibinfo {author}
  {\bibfnamefont {N.}~\bibnamefont {Van~Horne}}, \bibinfo {author}
  {\bibfnamefont {K.}~\bibnamefont {Bouzehouane}}, \bibinfo {author}
  {\bibfnamefont {K.}~\bibnamefont {Garcia}}, \bibinfo {author} {\bibfnamefont
  {C.}~\bibnamefont {Deranlot}}, \bibinfo {author} {\bibfnamefont
  {P.}~\bibnamefont {Warnicke}}, \bibinfo {author} {\bibfnamefont
  {P.}~\bibnamefont {Wohlhuter}}, \bibinfo {author} {\bibfnamefont {J.~M.}\
  \bibnamefont {George}}, \bibinfo {author} {\bibfnamefont {M.}~\bibnamefont
  {Weigand}}, \bibinfo {author} {\bibfnamefont {J.}~\bibnamefont {Raabe}},
  \bibinfo {author} {\bibfnamefont {V.}~\bibnamefont {Cros}}, \ and\ \bibinfo
  {author} {\bibfnamefont {A.}~\bibnamefont {Fert}},\ }\href {\doibase
  {10.1038/nnano.2015.313}} {\bibfield  {journal} {\bibinfo  {journal} {Nature
  Nanotechn.}\ }\textbf {\bibinfo {volume} {{11}}},\ \bibinfo {pages} {{444}}
  (\bibinfo {year} {{2016}})}\BibitemShut {NoStop}%
\bibitem [{\citenamefont {Chen}\ \emph {et~al.}(2013)\citenamefont {Chen},
  \citenamefont {Ma}, \citenamefont {N'Diaye}, \citenamefont {Kwon},
  \citenamefont {Won}, \citenamefont {Wu},\ and\ \citenamefont
  {Schmid}}]{Chen2013}%
  \BibitemOpen
  \bibfield  {author} {\bibinfo {author} {\bibfnamefont {G.}~\bibnamefont
  {Chen}}, \bibinfo {author} {\bibfnamefont {T.}~\bibnamefont {Ma}}, \bibinfo
  {author} {\bibfnamefont {A.~T.}\ \bibnamefont {N'Diaye}}, \bibinfo {author}
  {\bibfnamefont {H.}~\bibnamefont {Kwon}}, \bibinfo {author} {\bibfnamefont
  {C.}~\bibnamefont {Won}}, \bibinfo {author} {\bibfnamefont {Y.}~\bibnamefont
  {Wu}}, \ and\ \bibinfo {author} {\bibfnamefont {A.~K.}\ \bibnamefont
  {Schmid}},\ }\href@noop {} {\bibfield  {journal} {\bibinfo  {journal} {Nat.
  Commun.}\ }\textbf {\bibinfo {volume} {4}},\ \bibinfo {pages} {2671}
  (\bibinfo {year} {2013})}\BibitemShut {NoStop}%
\bibitem [{\citenamefont {Hrabec}\ \emph {et~al.}(2014)\citenamefont {Hrabec},
  \citenamefont {Porter}, \citenamefont {Wells}, \citenamefont {Benitez},
  \citenamefont {Burnell}, \citenamefont {McVitie}, \citenamefont {McGrouther},
  \citenamefont {Moore},\ and\ \citenamefont {Marrows}}]{Hrabec2014}%
  \BibitemOpen
  \bibfield  {author} {\bibinfo {author} {\bibfnamefont {A.}~\bibnamefont
  {Hrabec}}, \bibinfo {author} {\bibfnamefont {N.~A.}\ \bibnamefont {Porter}},
  \bibinfo {author} {\bibfnamefont {A.}~\bibnamefont {Wells}}, \bibinfo
  {author} {\bibfnamefont {M.~J.}\ \bibnamefont {Benitez}}, \bibinfo {author}
  {\bibfnamefont {G.}~\bibnamefont {Burnell}}, \bibinfo {author} {\bibfnamefont
  {S.}~\bibnamefont {McVitie}}, \bibinfo {author} {\bibfnamefont
  {D.}~\bibnamefont {McGrouther}}, \bibinfo {author} {\bibfnamefont {T.~A.}\
  \bibnamefont {Moore}}, \ and\ \bibinfo {author} {\bibfnamefont {C.~H.}\
  \bibnamefont {Marrows}},\ }\href {\doibase 10.1103/PhysRevB.90.020402}
  {\bibfield  {journal} {\bibinfo  {journal} {Phys. Rev. B}\ }\textbf {\bibinfo
  {volume} {90}},\ \bibinfo {pages} {020402} (\bibinfo {year}
  {2014})}\BibitemShut {NoStop}%
\bibitem [{\citenamefont {Han}\ \emph {et~al.}(2016)\citenamefont {Han},
  \citenamefont {Kim}, \citenamefont {Kim}, \citenamefont {Yin}, \citenamefont
  {Koo}, \citenamefont {Cho}, \citenamefont {Lee}, \citenamefont {Kläui},
  \citenamefont {Swagten}, \citenamefont {Koopmans},\ and\ \citenamefont
  {You}}]{Han2016}%
  \BibitemOpen
  \bibfield  {author} {\bibinfo {author} {\bibfnamefont {D.-S.}\ \bibnamefont
  {Han}}, \bibinfo {author} {\bibfnamefont {N.-H.}\ \bibnamefont {Kim}},
  \bibinfo {author} {\bibfnamefont {J.-S.}\ \bibnamefont {Kim}}, \bibinfo
  {author} {\bibfnamefont {Y.}~\bibnamefont {Yin}}, \bibinfo {author}
  {\bibfnamefont {J.-W.}\ \bibnamefont {Koo}}, \bibinfo {author} {\bibfnamefont
  {J.}~\bibnamefont {Cho}}, \bibinfo {author} {\bibfnamefont {S.}~\bibnamefont
  {Lee}}, \bibinfo {author} {\bibfnamefont {M.}~\bibnamefont {Kläui}}, \bibinfo
  {author} {\bibfnamefont {H.~J.~M.}\ \bibnamefont {Swagten}}, \bibinfo
  {author} {\bibfnamefont {B.}~\bibnamefont {Koopmans}}, \ and\ \bibinfo
  {author} {\bibfnamefont {C.-Y.}\ \bibnamefont {You}},\ }\href {\doibase
  10.1021/acs.nanolett.6b01593} {\bibfield  {journal} {\bibinfo  {journal}
  {Nano Letters}\ }\textbf {\bibinfo {volume} {16}},\ \bibinfo {pages} {4438}
  (\bibinfo {year} {2016})}\BibitemShut {NoStop}%
\bibitem [{\citenamefont {Je}\ \emph {et~al.}(2013)\citenamefont {Je},
  \citenamefont {Kim}, \citenamefont {Yoo}, \citenamefont {Min}, \citenamefont
  {Lee},\ and\ \citenamefont {Choe}}]{Je2013}%
  \BibitemOpen
  \bibfield  {author} {\bibinfo {author} {\bibfnamefont {S.-G.}\ \bibnamefont
  {Je}}, \bibinfo {author} {\bibfnamefont {D.-H.}\ \bibnamefont {Kim}},
  \bibinfo {author} {\bibfnamefont {S.-C.}\ \bibnamefont {Yoo}}, \bibinfo
  {author} {\bibfnamefont {B.-C.}\ \bibnamefont {Min}}, \bibinfo {author}
  {\bibfnamefont {K.-J.}\ \bibnamefont {Lee}}, \ and\ \bibinfo {author}
  {\bibfnamefont {S.-B.}\ \bibnamefont {Choe}},\ }\href {\doibase
  10.1103/PhysRevB.88.214401} {\bibfield  {journal} {\bibinfo  {journal} {Phys.
  Rev. B}\ }\textbf {\bibinfo {volume} {88}},\ \bibinfo {pages} {214401}
  (\bibinfo {year} {2013})}\BibitemShut {NoStop}%
\bibitem [{\citenamefont {Va\v{n}atka}\ \emph {et~al.}(2015)\citenamefont
  {Va\v{n}atka}, \citenamefont {Rojas-S\'{a}nchez}, \citenamefont {Vogel},
  \citenamefont {Bonfim}, \citenamefont {Thiaville},\ and\ \citenamefont
  {Pizzini}}]{Vanatka2015}%
  \BibitemOpen
  \bibfield  {author} {\bibinfo {author} {\bibfnamefont {M.}~\bibnamefont
  {Va\v{n}atka}}, \bibinfo {author} {\bibfnamefont {J.-C.}\ \bibnamefont
  {Rojas-S\'{a}nchez}}, \bibinfo {author} {\bibfnamefont {J.}~\bibnamefont
  {Vogel}}, \bibinfo {author} {\bibfnamefont {M.}~\bibnamefont {Bonfim}},
  \bibinfo {author} {\bibfnamefont {A.}~\bibnamefont {Thiaville}}, \ and\
  \bibinfo {author} {\bibfnamefont {S.}~\bibnamefont {Pizzini}},\ }\href@noop
  {} {\bibfield  {journal} {\bibinfo  {journal} {J. Phys.: Condens. Matter}\
  }\textbf {\bibinfo {volume} {27}},\ \bibinfo {pages} {32002} (\bibinfo {year}
  {2015})}\BibitemShut {NoStop}%
\bibitem [{\citenamefont {Pham}\ \emph {et~al.}(2016)\citenamefont {Pham},
  \citenamefont {Vogel}, \citenamefont {Sampaio}, \citenamefont {Va\v{n}atka},
  \citenamefont {Rojas-Sanchez}, \citenamefont {Bonfim}, \citenamefont
  {Chaves}, \citenamefont {Choueikani}, \citenamefont {Ohresser}, \citenamefont
  {Otero}, \citenamefont {Thiaville},\ and\ \citenamefont
  {Pizzini}}]{Pham2016}%
  \BibitemOpen
  \bibfield  {author} {\bibinfo {author} {\bibfnamefont {T.~H.}\ \bibnamefont
  {Pham}}, \bibinfo {author} {\bibfnamefont {J.}~\bibnamefont {Vogel}},
  \bibinfo {author} {\bibfnamefont {J.}~\bibnamefont {Sampaio}}, \bibinfo
  {author} {\bibfnamefont {M.}~\bibnamefont {Va\v{n}atka}}, \bibinfo {author}
  {\bibfnamefont {J.-C.}\ \bibnamefont {Rojas-Sanchez}}, \bibinfo {author}
  {\bibfnamefont {M.}~\bibnamefont {Bonfim}}, \bibinfo {author} {\bibfnamefont
  {D.~S.}\ \bibnamefont {Chaves}}, \bibinfo {author} {\bibfnamefont
  {F.}~\bibnamefont {Choueikani}}, \bibinfo {author} {\bibfnamefont
  {P.}~\bibnamefont {Ohresser}}, \bibinfo {author} {\bibfnamefont
  {E.}~\bibnamefont {Otero}}, \bibinfo {author} {\bibfnamefont
  {A.}~\bibnamefont {Thiaville}}, \ and\ \bibinfo {author} {\bibfnamefont
  {S.}~\bibnamefont {Pizzini}},\ }\href@noop {} {\bibfield  {journal} {\bibinfo
   {journal} {{EPL}}\ }\textbf {\bibinfo {volume} {{113}}},\ \bibinfo {pages}
  {67001} (\bibinfo {year} {{2016}})}\BibitemShut {NoStop}%
\bibitem [{\citenamefont {Chaves}\ \emph {et~al.}(2017)\citenamefont {Chaves},
  \citenamefont {Ajejas}, \citenamefont {Krizakova}, \citenamefont {Vogel},\
  and\ \citenamefont {Pizzini}}]{Chaves2017}%
  \BibitemOpen
  \bibfield  {author} {\bibinfo {author} {\bibfnamefont {D.}~\bibnamefont
  {Chaves}}, \bibinfo {author} {\bibfnamefont {F.}~\bibnamefont {Ajejas}},
  \bibinfo {author} {\bibfnamefont {V.}~\bibnamefont {Krizakova}}, \bibinfo
  {author} {\bibfnamefont {J.}~\bibnamefont {Vogel}}, \ and\ \bibinfo {author}
  {\bibfnamefont {S.}~\bibnamefont {Pizzini}},\ }\href@noop {} {\bibfield
  {journal} {\bibinfo  {journal} {arXiv:1708.08516}\ } (\bibinfo {year}
  {2017})}\BibitemShut {NoStop}%
\end{thebibliography}
\end{document}